\documentclass[twocolumn]{jpsj3}
\usepackage{color}

\def\runauthor{S. \textsc{Mao} et al.}

\title{
Analytic Theory of Edge Modes in Topological Insulators}

\author{ Shijun Mao$^{1,2}$, Yoshio Kuramoto$^{1}$, Ken-Ichiro  Imura$^{1,3}$, and Ai  Yamakage$^{1}$ }
\inst {Department of Physics, Tohoku University,
Sendai 980-8578, Japan $^1$\\Department of Physics, Tsinghua
University, Beijing 100084, P.R.China $^2$\\Department of Quantum
Matter, AdSM, Hiroshima University, Higashi-Hiroshima 739-8530,
Japan $^3$}

\recdate{\today}

\abst{
Spectrum and wave function of gapless
edge modes are derived analytically for a tight-binding model of topological
insulators on square lattice.
Particular attention is paid to dependence on edge geometries such as the straight (1,0) and zigzag (1,1) edges in the thermodynamic limit.
The key technique is to identify operators
that combine to annihilate the edge state in the effective one-dimensional (1D) model with momentum along the edge.
In the (1,0) edge, the edge mode is present either around the center of 1D Brillouin zone
or its boundary,
depending on location of the bulk excitation gap.
In the (1,1) edge, the edge mode is always present both at the center and
near the boundary.
Depending on system parameters, however,
the mode is absent in the middle of the Brillouin zone.
In this case the binding energy of the edge mode near the boundary
is extremely small;  about $10^{-3}$ of the overall energy scale.
Origin of this minute energy scale is discussed.
}

\kword{zigzag lattice, effective Hamiltonian, flat edge spectrum,
reentrant edge mode }

\begin{document}
\maketitle

\section {Introduction}
Recently,
a new class of topological insulator (TI), also referred to as quantum spin Hall (QSH) insulator, has been attracting much interest in both theory \cite{{he1}, {km1}, {km2}, {km3},
{zhang1}, {BHZ}, {zhang4}} and experiments \cite{{ex1}, {ex2}, {ex3}}.
Different from conventional insulators, QSH
insulators have topologically protected helical edge states
with the spectrum lying in the bulk insulating gap.\cite{{he1}, {he2}, {he3}}
Two-dimensinal (2D) topological insulator has been realized in HgTe/CdTe
quantum wells\cite{ex1} following theoretical suggestion of 
Bernevig, Hughes and Zhang\cite{BHZ} who proposed a useful model, hereafter referred to as the BHZ Model.
Helical edge state has been further studied using the resultant continuum
model.\cite{{cm}, {cm2}}
The BHZ model is not realistic away from the zone center,
since it is based on the $\mib k \cdot \mib p$
perturbation theory and envelope-function approximation.
However, 
a regularization of the model using the tight-binding scheme
has the simple structure for the whole Brillouin zone (BZ), and
is suitable for studying the general property of QSH systems. 
Note that characterization of the topological property of the system
requires information of wave functions over the whole BZ.

In this paper, we derive the spectrum of edge modes analytically taking the tight-binding version of the BHZ model for the square lattice. In
contrast to previous study of edge modes,\cite{HgTe,cm} we take not
only the straight (1,0) edge but also the zigzag (1,1) edge, and
make detailed comparison of respective edge modes. This comparison
is partially inspired by the remarkable difference between the
zigzag and armchair edges in graphene \cite{{G1},{G2}}.
In contrast to the latter, however, the edges in the present model should not be taken as representing the HgTe/CdTe quantum wells since the short-distance behavior of the model is not realistic.  Nevertheless, the intuition gained by the exact solution of the simplified model should provide useful information for understanding more complicated systems.

In a separate paper\cite{imura},
we have already derived the spectrum of both straight and zigzag edges relying partially on numerical method.
In particular, we have found
for the zigzag edge a reentrant mode with tiny binding energy.
Since the relevant energy is so small as compared with the overall energy, it is desirable to characterize its nature analytically.
Then the origin of tiny binding energy should be clarified.
In the present paper, we derive fully analytic expressions of not only the spectrum for both edges, but also momentum regions allowed for the modes.
Namely we derive critical momentum where the edge mode merges with bulk excitations.
Our analytical method is a systematic generalization of previous ones
\cite{Ed1, HgTe} so that the spectrum can be obtained for general direction of the edge.

This paper is organized as follows:
In \S 2,  we review the lattice version of the
BHZ model with nearest-neighbor transfer, paying attention to its symmetry property
in the 2D BZ.
Sections 3 and 4 are devoted to analytic derivation of edge modes for
straight and zigzag edges, respectively.
We use a systematic method to derive the spectrum
in terms of the ``annihilator".
Existent regions of edge modes are derived for both straight and
zigzag edges. In the zigzag case, we find a novel reentrant edge
mode with a tiny binding energy below the bulk spectrum. Finally, we
summarize the results and discuss their implication in Sec. 5.

\section{Model with particle-hole symmetry}

We consider the BHZ model given by the following $4\times 4$ matrix:
\begin{align}
&H (\mib k)=\left[
\begin{array}{cc}
h(\mib k)  &   0 \\
0  &   h^*(-\mib k) \\
\end{array}
\right], \label{Htot}
\end{align}
where $\mib k=(k_x,k_y)$ is a 2D crystal momentum, measured from
$\Gamma$-point. The lower-right
block $h^* (-\mib k)$ is a $2\times 2$ matrix, and
is deduced
from the upper-left block
$h(\mib k)$ by time reversal transformation.
$h(\mib k)$ is
parametrized as,
\begin{equation}
h (\mib k) = \mib d (\mib k) \cdot \mib \sigma = \left[
\begin{array}{cc}
d_z  &  d_x -i d_y  \\
d_x  +i d_y &  - d_z
\end{array}
\right], \label{hk}
\end{equation}
where $d_{i}(\mib k)$ are given by
\begin{align}
d_x ({\mib k})  = A\sin k_x, \quad
d_y ({\mib k})  = A\sin k_y, \\
d_z ({\mib k})  = \Delta -2B (2-\cos k_x-\cos k_y),
\end{align}
with the lattice constant $a$ set to unity\cite{BHZ}. 
We only
consider the case $A>0,\ B>0$ and $\Delta \ge 0$ in this paper. The
same signs of $B$ and $\Delta$ are necessary for topological
insulator. 
The solution for $A<0$ is trivially obtained from that 
for $H(-\mib k)$.
Each row and column of eq. (\ref{hk})
 represent spin-orbital states associated with
the $s$-type $\Gamma_6$ and the $p$-type $\Gamma_8$ orbitals of the
3D band structure of HgTe and CdTe.
Other parameters
which appear in Ref.\citen{BHZ}, i.e., $C$ and $D$ have been set to
zero.
As a result, the spectrum has the particle-hole symmetry that simplifies the analysis.

The $2\times 2$-matrix Hamiltonian $h(\mib k)$ is equivalent to the
following tight-binding Hamiltonian on the square lattice:
\begin{align}
h_{\uparrow} &= \sum_{I,J}  c_{I,J}^\dagger \hat{\cal E} c_{I,J}
\nonumber\\
&+\left( c_{I,J}^\dagger \hat{t}_x c_{I+1,J} + c_{I,J}^\dagger
\hat{t}_y c_{I,J+1}
 + h.c.\right), \label{hBHZ}
\end{align}
where we have introduced for each site $(I,J)$ the two-component
field:
\begin{align}
c^\dagger = ( c^\dagger_{s \uparrow}, c^\dagger_{p \uparrow} )
\end{align} with orbitals $s,p$.  The tight-binding parameters are given by
\begin{align}
\hat{\cal E} &=(\Delta-4B)\sigma_z,
\label{cal_E} \\
\hat{t}_x&= -i {A\over 2}\sigma_x + B \sigma_z,
\\
\hat{t}_y &= -i {A\over 2}\sigma_y + B \sigma_z, \label{txy}
\end{align}
each of which is a $2\times 2$ matrix. The down spin part
$h_\downarrow$ corresponding to the lower-right block of
eq.(\ref{Htot}) is obtained by replacing $\hat{t}_{i}$ with its
complex conjugation $\hat{t}_{i}^*$.

We consider the BHZ model over the whole BZ of the
square lattice. Bulk energy $E_{\rm b}({\mib k})$
is written as
\begin{align}
E_{\rm b}({\mib k})
&= \pm \left\{
A^2\left( \sin^2 k_x+\sin^2 k_y \right)\right.
\nonumber \\
&+\left.
\left[ \Delta-4B+2B\left(
\cos k_x+\cos k_y \right) \right]^2 \right\}^{1/2},
 \label{bulk}
\end{align}
which is symmetric with respect to positive (conduction) and negative (valence) energy bands,
being the signature of the particle-hole symmetry.
As one varies mass parameter $\Delta$, there appear four gap closing
points, namely at $\Gamma=(0,0)$, $X=(\pi,0)$, $X'=(0,\pi)$ and
$M=(\pi,\pi)$.
Note that these points are invariant against time-reversal operation.
Gap closing at $\Gamma$ occurs when $\Delta=0$,
whereas the gap closing at $X$ and $X'$ occurs simultaneously when
$\Delta=4B$, at $M$ when $\Delta=8B$.

\section{Straight edge}
\begin{figure}
\begin{center}
\includegraphics[width=0.8\linewidth]{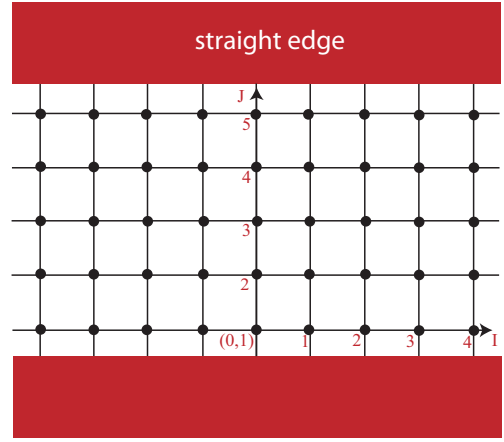}
\caption{Straight edge lattice ribbon with two boundaries in (1,0)
direction.} \label{fig1}
\end{center}
\end{figure}
%
\subsection{Effective one-dimensional form}
Let us first consider the
geometry (see Fig. \ref{fig1}), in which electrons are confined to
$N_{r}$ rows in a strip between $y=1$ and $y=N_{r}$, i.e., the two
edges are along $x$-axis.
Translational invariance along $x$-axis
allows for constructing a 1D Bloch state with a crystal momentum
$k=k_x$:
\begin{equation}
|k,J\rangle=\frac 1{\sqrt {N_c}} \sum_I e^{ikI} |I,J\rangle,
\label{Bloch}
\end{equation}
where $I, J$ represents lattice site,
$N_c$ is the number of sites along the $x$-axis, and
$|I,J\rangle=c_{I,J}^\dagger |0\rangle$.
In order to introduce the
edges, it is convenient to rewrite eq. (\ref{hBHZ}) in form of a
hopping Hamiltonian between neighboring rows. In terms of the
two-component creation and annihilation operators $c^\dagger_{J}(k)$
and $c_{J}(k)$ associated with Bloch state (\ref{Bloch}), one can
rewrite eq. (\ref{hBHZ}) as
\begin{align}
h_{\uparrow} &=\sum_{k} h_{01}(k), \\
h_{01}(k) &= \sum_{J} c^\dagger_J(k) \hat{\cal E} (k) c_J(k)
\nonumber \\& + \sum_{J} \left[ c^\dagger_{J}(k) \hat{t}_y
c_{J+1}(k) + h.c. \right] \label{hNR}
\end{align}
where $\hat{\cal E} (k)$ is given by
\begin{equation}
\hat{\cal E} (k) =A \sin k \sigma_x + \left( \Delta_B +2\cos k \right)
\sigma_z, \label{epsilon}
\end{equation}
Here and in the following,
we use the notation $\Delta_B \equiv \Delta-4B$, and
take the energy unit so that $B=1$.
The corresponding Schr\"{o}dinger equation is given by
\begin{equation}
\hat{\cal E}(k)\Psi_J +\hat t_y \Psi_{J-1}+\hat t _y^\dagger
\Psi_{J+1} = E_\uparrow (k)\Psi_J \label{TB2}
\end{equation}
where $\Psi_J$ is the two-component amplitude with row index $J$.
The straight edges along the $J=1$ row and $J=N_{r}$ row can be
implemented by open boundary condition $\Psi_0=\Psi_{N_r+1}=0$.

It has been shown \cite{cm} in the continuum approximation of
eq.(\ref{TB2}) that there appear modes, which are localized on both
edges of the system, with the particle-hole symmetric spectrum and
a minimum gap at $k=0$.  The origin of the gap is the overlap of
wave functions on different edges. As $N_r$ tends to infinity, the
gap disappears since the overlap becomes zero.  In this limit one
can identify edge modes localized on either edge. In the thermodynamic
limit,  the spectrum $E(k)$ of an edge mode near $J=1$ becomes an
odd function of $k$, and the other edge mode has the spectrum
$-E(k)$. The continuum approximation of Ref.\citen{cm} has a limited
validity, but emergence of the zero mode at $k=0$ in the
thermodynamic limit $N_r\rightarrow \infty$ is guaranteed by the
time-reversal and particle-hole symmetries \cite{Ed1,HgTe}. In the
following, we deal with the thermodynamic limit.

\subsection{Separation into Hermitian and annihilating parts}

Our strategy to obtain the edge spectrum is best illustrated in the straight edge.
Although the spectrum in this case has already been obtained in the literature \cite{Ed1,HgTe},
we here present our way of derivation that can be extended to the zigzag edge.
%
We try an edge state solution with property: $\Psi_{J+1} =
\lambda\Psi_J\equiv \lambda^{J+1}\Psi$ with
$|\lambda|<1$\cite{{Ed1},{Ed2}},
and derive the vector $\Psi$.
Then eq.(\ref{TB2}) can be written
in the following form:
\begin{equation}
\left[  \hat{\cal E}(k) +\lambda \hat t _y^\dagger +\frac 1\lambda
\hat t_y \right] \Psi \equiv P_{01}(\lambda, k)\Psi =E_\uparrow (k)\Psi,
\label{TB3}
\end{equation}
where $ P_{01}(\lambda, k) $ can be rearranged as
\begin{align}
P_{01}(\lambda, k) = \mib \gamma\cdot \mib \sigma,
\label{P01}
\end{align}
with components
\begin{align}
\gamma_x &= A\sin k
\label{gammax}
\end{align}
\begin{align}
\gamma_y &=  i \frac A2  \left( \lambda- \lambda^{-1} \right) 
\label{gammay}
\end{align}
\begin{align}
\gamma_z &=  \left( \lambda+ \lambda^{-1} \right) +\Delta_B+2\cos k
\label{gammaz}
\end{align}
For a general complex vector $\mib \gamma$,
the eigenvalues of $P_{01}$ are also complex.
However, we obtain real energy $E_\uparrow(k)$ in eq.(\ref{TB3})
if one of the following conditions is met: \\
(a) all components of $\mib \gamma $ are real (including zero); \\
(b) the nonzero complex components combine to give zero when acting on the edge state.  Such combination of operators is referred to as
annihilator.
\\
The condition (a) is not relevant here, because we would then have two eigenvalues $\pm E_\uparrow (k)$ and corresponding two eigenfunctions
for a given $k$.   Actually we should have only one edge mode for $h_\uparrow (k)$.
Hence we have to accept the condition (b), and separate $P_{01}$
into a Hermitian part that gives the spectrum by diagonalization,
and the rest that makes up the annihilator.

From eq.{(\ref{gammay}) and (\ref{gammaz})}, coefficients $\gamma_y$ and $\gamma_z$ are both real only if $|\lambda|=1$.  On the other hand,
due to time-reversal symmetry and particle-hole symmetry of the system, the eigenvalue $E_\uparrow(k)$ is an odd function of $k$ in the thermodynamic limit \cite{HgTe,cm}.
Therefore, $\sigma_y$ and 
$\sigma_z$ must both belong to the annihilator, and 
the only component to be diagonalized is $\sigma_x$. The
annihilator corresponds to either $\sigma_y+i\sigma_z$ or
$\sigma_y-i\sigma_z$,
depending on the eigenvalue $\pm 1$ of $\sigma_x$.


Accordingly, we decompose $P_{01}(\lambda, k)$ as
\begin{align}
&P_{01}(\lambda, k)=H_{01}+F_{01}, \\
&H_{01}=A\sin k \, \sigma_x, \\
&F_{01}=\gamma_y \sigma_y+ \gamma_z \sigma_z,
\end{align}
where
$F_{01}$ should form the annihilator.
Namely,
we impose the relation $\gamma_z= s i\gamma_y$, according to the eigenvalue $s=\pm 1$ of $\sigma_x$.
The relation is equivalent to
\begin{equation}
(1+\frac{sA}{2})\lambda+(1-\frac{sA}{2})\lambda^{-1}+m_k=0,
\label{lambda01}
\end{equation}
where we have introduced the notation:
\begin{align}
m_k=\Delta_B+2 \cos k.
\end{align}
By solving eq.(\ref{lambda01}) with $A\neq 2$, we obtain
\begin{equation}
\lambda_{\pm} (s)=\frac{-m_k\pm\sqrt{m^2_k+A^2-4}}{2+sA},
\label{lambda10_pm}
\end{equation}
with the relation
\begin{align}
\lambda_\pm (-s) = 1/\lambda_\mp (s).
\label{recip}
\end{align}
Provided the eigenvalue equation
\begin{align}
H_{01}\Psi &= E_\uparrow(k)\Psi. \label{H101}
\end{align}
is satisfied,
then equality $F_{01}\Psi=0$ follows with eq.(\ref{lambda01}).
In this way
the edge spectrum is simply
derived as
\begin{equation}
E_\uparrow(k)=sA\sin k.
\label{E01}
\end{equation}
with the eigenstate written as
\begin{align}
\Psi (s) =
\begin{pmatrix} 
1\\s
\end{pmatrix}.
\end{align}
Note that only one of $s=\pm 1$ is relevant, as discussed below.

The boundary condition $\Psi_{J=0}=0$
requires the edge
state to have the form
\begin{align}
\Psi_J (s) =
\left[
\lambda_{+}(s)^{J}-\lambda_{-}(s)^{J} \right]
\Psi (s),
\end{align}
apart from the normalization factor.
Since the wave function should decay as $J$ increases,
the edge state can only be realized with
$|\lambda_\pm (s)|<1$.
Because of the relation eq.(\ref{recip}), there is at most one $s$ for given $k$ that describes the edge mode with the spectrum eq.(\ref{E01}).
Hence there is only one edge state per spin and per momentum.

In a similar manner, we obtain another edge mode for
$h_{\downarrow}$.
The corresponding energy $E_\downarrow (k)$ is given by
\begin{align}
E_\downarrow (k) = E_\uparrow(-k) = -E_\uparrow(k),
\end{align}
where the first equality corresponds to the time-reversal symmetry.
The particle-hole symmetry connecting the rightmost and leftmost sides
with the same $k$ involves different spins.
Note that the Kramers pair has the same $\lambda_\pm(s)$ for given $k$ and $-k$.
In certain range of $k$, however, the edge modes do not exist.
This problem is studied in detail in the next section.

\subsection{Allowed momentum range for edge modes}

A pair of gapless edge modes per edge are always present in the topological insulator phase (TI) that appears for $0<\Delta<8$, {\it i.e.}, $|\Delta_B|<4$ in the BHZ model.
However,
the number of zero points of helical edge states should be an odd multiple of two, including the degeneracy,  in the 1D BZ \cite{he2}.
If the edge modes with the spectrum $E=\pm A\sin k$ were present for the whole BZ, the zero points amount to 
four (an even multiple of two)
which violates the topological stability \cite{nogo1}.
Hence the edge modes must merge into
bulk excitations at finite $\kappa_m$, and degenerate zero points occur either at $k=0$ or $k=\pi$, but not at both.
Note that
$k=0$ and $k=\pi$ are two time reversal invariant momenta in
1D BZ.

\begin{figure}
\begin{center}
\includegraphics[width=0.8\linewidth]{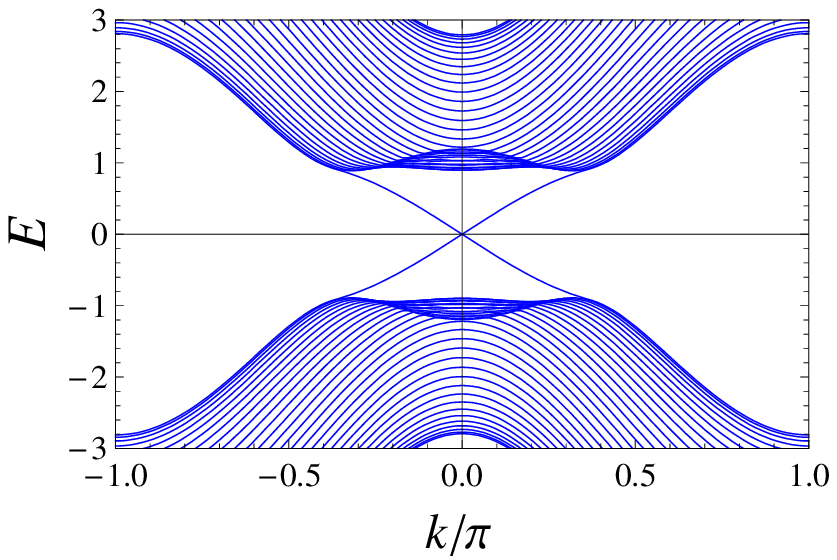}\\
\includegraphics[width=0.8\linewidth]{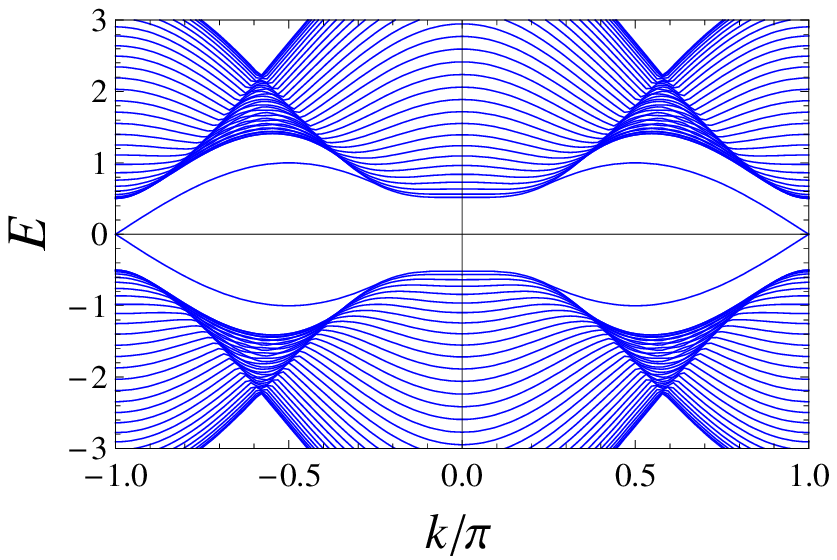}
\caption{1D energy bands in the straight edge: TI-1
(top) with $\Delta=1.2$ and TI-2 (bottom) with $\Delta=4.5$.
In both cases, we set $A=1$.
The edge modes have the spectrum $\pm A\sin k$ in the existent range of $k$.
}
\label{fig2}
\end{center}
\end{figure}

With $\Delta_B=0$, the energy gap closes at X points $(\pi,0)$ and $(0,\pi)$
in the 2D BZ, as seen from eq.(\ref{bulk}).
Let us classify the case of $-4<\Delta_B<0$ as TI-1, and
the case of $0<\Delta_B<4$ as TI-2.
Figure \ref{fig2} shows the spectrum of edge modes in the TI-1 (upper panel) and TI-2 (lower panel).
In the TI-1, the edge modes intersect at $k=0$, whereas in TI-2 they meet at $k=\pi$.
Also shown is approximate bulk spectrum in the system that is derived from  $E_{\rm b}(k, k_y)$ by fixing $k_y$ to $2\pi n/N_r$ with $n=1,\ldots,N_r$.  Each curve for the bulk spectrum corresponds to integer $n$.

We now derive the pair $\pm k_m$ of momentum where the edge mode merges with bulk excitations.
In the following, we always assume $k\ge 0$ for simplicity.
The merging occurs when the larger of $|\lambda_\pm(s)|$ becomes unity.  Note that such $\lambda_\pm(s)$ is real, since otherwise
$|\lambda_\pm(s)|$ is independent of $m_k$, and hence of $k$.
For complex $\lambda_\pm(s)$ we obtain from eq.(\ref{lambda10_pm}),
\begin{align}
|\lambda_\pm(s)|^2 = \frac{2-s A}{2+s A},
\end{align}
which is less than unity only with $s=1$.
In this section we deal with the case $0<A<2$ that allows complex
$\lambda_\pm(s)$.
Then, edge modes 
are assured to be present for such $k$ with complex $\lambda_\pm(1)$,
which is simply written as $\lambda_\pm$ hereafter.
Edge modes in the case of $A>2$ will be discussed in \S 5.

\subsubsection*{TI-1 regime}
In this case gapless points are present at $k=0$.
Let us assume $m_k=\Delta_B+2 \cos k<0$ when merging occurs at
$k=k_{m1}$, 
Then we obtain $|\lambda_{+}|>|\lambda_{-}|$ for real $\lambda_{\pm}$, and  
the condition for merging is reduced to
\begin{align}
\lambda_{+} =\frac{-m_k +\sqrt{m^2_k+A^2-4}}{2+A}=1.
\end{align}
from eq.(\ref{lambda10_pm}).
Then we obtain
\begin{equation}
\cos k_{m1} = 1-\Delta/2, \label{k_m}
\end{equation}
which justifies the assumption $m_k<0$.
One can check that there is no solution if we assume $m_k>0$.
We note that eq.(\ref{k_m}) has already been obtained by K\"{o}nig et al.\cite{HgTe}

The condition for merging is also to
have the same energy as the lowest bulk excitation 
for given $k$.
The minimum of 
$E_b(k,k_y)$ 
may occur either at $k_y=0$ or 
$k_y=\pm \arccos \left[ 2m_{k}/(A^2-4) \right]$ 
depending on the value of $k$.
If the threshold of bulk excitations with $k=k_{m1}$ occurs at
$k_y=0$,
merging momentum $k_{m1}$ is simply obtained from
eq.(\ref{bulk}) as
\begin{equation}
\Delta_B +2+2\cos k = m_k+2=0,
\end{equation}
which is consistent with eq.(\ref{k_m}).
Namely, the condition $\lambda_+=1$ is equivalent to having the same energy for edge mode and for the minimum of bulk excitations.
Furthermore, it is easily seen that the group velocity $A\cos k_{m1}$ at  merging point is common to both edge mode and the lowest bulk excitation.  Namely, the edge mode vanishes at such $k$ that it has the common tangent with the threshold of
bulk excitations $E_b(k,k_y)$ with $k_y=0$.

The crossings of 1D energy bands in Fig.\ref{fig2} indicates the transition from the lowest bulk excitation $E_b(k,k_y)$ with $k_{y}=0$ to with 
$k_y=\pm \arccos \left[ 2m_{k}/(A^2-4) \right]$. 
The critical momentum $k_{c1}$ satisfies
the condition
\begin{equation}
\partial^2 E_{\rm b} (k_{c1}, k_y)/\partial k_y^2\vert_{k_y=0}=0,
\end{equation}
which gives the solution
\begin{equation}
\cos k_{c1} = 1-\frac \Delta{2}+\frac{A^2}{4}. \label{coskc}
\end{equation}
The threshold has $k_y=0$ for $k>k_{c1}>0$.
By comparing with eq.(\ref{k_m}), we find $\cos k_{c1}>\cos k_{m1}$, which
means $0\le k_{c1}< k_{m1}$.
Namely,  merging with bulk excitations indeed occurs in the range where $k_y=0$ corresponds to the threshold.

\subsubsection*{TI-2 regime}
In this case edge modes exist around $k=\pi$.  Let us assume $m_k>0$ when merging occurs at $k=k_{m2}>0$.
The condition for  merging is now given by
\begin{align}
\lambda_{-} =\frac{-m_k -\sqrt{m^2_k+A^2-4}}{2+A}=-1,
\label{lambda-TI-1}
\end{align}
which gives $m_k=2$ as the solution, or
\begin{align}
\cos k_{m2} = 1-\Delta_B/2  = 3-\Delta/2.
\label{k_m2}
\end{align}
In the TI-2 regime,
the threshold of bulk excitations occurs at $k_{y}=\pi$ or 
$k_y=\pm \arccos \left[ 2m_{k}/(A^2-4) \right]$. 
It can be checked that the bulk energy at $\mib k= (k_{m2}, \pi)$ becomes the same as the edge mode with the condition (\ref{k_m2}).
Hence merging occurs with bulk excitations $E_b(k,k_y)$ with $k_y=\pi$.

The critical momentum $k=k_{c2}$, below which 
the minimum of $E_b(k,k_y)$ no longer occurs at $k_y=\pi$,
can be obtained by condition
\begin{equation}
\partial^2 E_{\rm b} (k_{c2}, k_y)/\partial k_y^2\vert_{k_y=\pi}=0,
\end{equation}
which gives the solution
\begin{equation}
\cos k_{c2} = 1-\frac {\Delta_B}{2}-\frac{A^2}{4}. \label{coskc2}
\end{equation}
Thus we have the relation $\cos k_{c2}<\cos k_{2m}$,
or $k_{c2} > k_{2m}$.
Hence merging indeed occurs in the range where $k_y=\pi$ corresponds to the threshold.
In this way, we have quantified important characteristics of the edge modes shown in Fig.\ref{fig2}.

\section{Zigzag edge}
\subsection{Effective one-dimensional form}
\begin{figure}
\begin{center}
\includegraphics[width=0.8\linewidth]{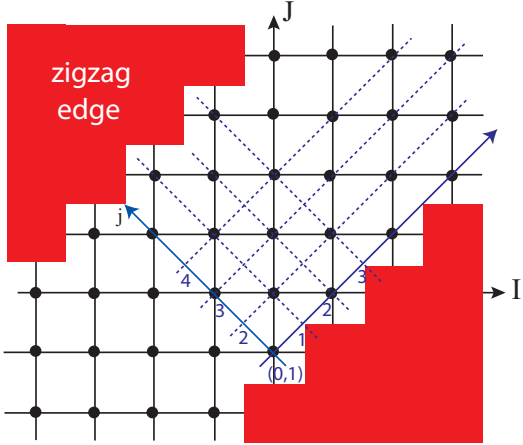}
\caption{Zigzag edge lattice ribbon, with two boundaries in (1,1) direction.}
\label{fig3}
\end{center}
\end{figure}

Let us now consider a zigzag edge geometry, as illustrated in Fig.\ref{fig3}.
 Electrons in a zigzag edge geometry are confined in
a diagonal strip: $1\le y-x \le N_{r}$, provided the edges are
placed at $y-x=1$ and $y-x=N_{r} $, normal to the
$(1,-1)$-direction. The translational invariance remains along the
$(1,1)$-direction, where the conserved momentum is given by
$p = (k_x+k_y)/\sqrt 2$. For
notational convenience we introduce $\kappa = p/\sqrt 2$ with
${-\pi/2<\kappa\le \pi/2}$ and define the new basis set as
\begin{equation}
|\kappa,j\rangle =\frac 1{\sqrt {N_c}} \sum_I \exp\left[ i\kappa \left( 2I+
j \right)\right]
 |I,I+j\rangle,
\label{Bloch11}
\end{equation}
where the site summation goes along the (1,1) direction. The phase
factor is so chosen that it becomes unity for the state $|-\!
J,J\rangle$. Then, the amplitude $\Phi_j$ in this basis satisfies
the Schr\"{o}dinger equation analogous to eq.(\ref{TB2}):
\begin{equation}
\hat{\cal E}\Phi_j +\hat t_{11}(\kappa) \Phi_{j-1}+\hat t^\dagger
_{11}(\kappa) \Phi_{j+1} = E_\uparrow (\kappa) \Phi_j \label{TB11}
\end{equation}
where $\hat{\cal E}$ has been defined by eq.(\ref{cal_E}) and
hopping matrix $\hat{t}_{11}(\kappa)$ is given by
\begin{align}
\hat{t}_{11}(\kappa) &
=\frac{i}{2}Ae^{-i\kappa}\sigma_x-\frac{i}{2}Ae^{i\kappa}\sigma_y+2\cos
k \sigma_z \label{gamma_zig}
\end{align}

We impose the boundary condition: $\Phi_0=\Phi_{N_r+1}=0$, which is
consistent with the zigzag edge geometry. Assuming eigenstate of
eq.(\ref{TB11}) with property
$\Phi_j=\lambda\Phi_{j-1}=\lambda^{j}\Phi$, where
$|\lambda|<1$,\cite{{Ed1},{Ed2}} we obtain
\begin{align}
\left (\hat{\cal E} +\lambda \hat
t^\dagger_{11}(\kappa)+\lambda^{-1} \hat t _{11}(\kappa)\right)\Phi
& =P_{11}(\lambda,\kappa)\Phi= E_\uparrow (\kappa) \Phi \label{TB112}
\end{align}

For later reference purpose, we write the bulk energy $E_{\rm b}$ in
terms of variables $\kappa = (k_x+k_y)/2$ and 
$\xi = (k_x-k_y)/2$.
From eq.(\ref{bulk}) we obtain
\begin{align}
E_{\rm b}(\kappa,\xi)
&= \pm \left[
2A^2\left( \sin^2 \kappa \cos^2\xi
+ \cos^2 \kappa \sin ^2\xi
\right)\right.
\nonumber \\
&+\left.
\left(
 \Delta_B+4
\cos \kappa \cos\xi  \right)
^2 \right]^{1/2}.
 \label{bulk11}
\end{align}

\subsection{Derivation of spectrum in thermodynamic limit}

We will separate $P_{11}$ into the Hermitian part
$H_{11}$ and the corresponding annihilator $F_{11}$.
The separation now is not straightforward in contrast with the case of straight edge.
As a preliminary, we introduce
the following matrices:
\begin{align}
\sigma_{X}&=\left( \sigma_x+\sigma_y \right)/\sqrt{2},\\
\sigma_{Y}&=\left( \sigma_y-\sigma_x \right)/\sqrt{2}.
\end{align}
Then we obtain
\begin{align}
&\hat{t}_{11}(\kappa) = \frac{A}{\sqrt{2}}\left(
\sin {\kappa}\, \sigma_{X} -i \cos {\kappa}\,\sigma_{Y}\right)+
2\cos {\kappa}\, \sigma_{z}. \label{t11_zig}
\end{align}
We note that the spectrum of each edge mode is an odd function of $\kappa$.  Then we introduce a variable $\theta$, which is an odd function of $\kappa$, and make the following transformation:
\begin{align}
\sigma_{\theta x}&={\cos {\theta}}\, \sigma_X + {\sin{\theta}}\, \sigma_z,
\label{sigma_mux}
\\
\sigma_{\theta z}&={\cos {\theta}}\, \sigma_z -{\sin{\theta}}\, \sigma_X,
\end{align}
and rewrite as $\sigma_{\theta y}=\sigma_Y$. 
They keep the commutation property:
\begin{align}
\left[ 
\sigma_{\theta x}, \sigma_{\theta y} \right] = 2i\sigma_{\theta z}, 
\end{align}
and analogous cyclic ones that are the same as the original Pauli matrices.
Then we obtain
\begin{align}
\hat{\cal E}&=\Delta_B \left(\sin{\theta\, }\sigma_{\theta x}+\cos{\theta\,
}\sigma_{\theta z} \right), \label{cal-E}
\\
\lambda \hat{t}^\dagger_{11}+\lambda^{-1} \hat{t}_{11} &=\gamma_{\theta
x}\sigma_{\theta x}+\gamma_{\theta y}\sigma_{\theta y} +\gamma_{\theta
z}\sigma_{\theta z},
\end{align}
where
\begin{align}
\gamma_{\theta x}&=(\lambda+\lambda^{-1})\left(
\frac{A}{\sqrt{2}}\sin{\kappa}\cos{\theta}+2\cos{\kappa}\sin{\theta} \right),  \\
\gamma_{\theta y}&=i
(\lambda-\lambda^{-1})\frac{A}{\sqrt{2}} \cos {\kappa}, \\
\gamma_{\theta z} &= (\lambda+\lambda^{-1}) \left( 2\cos{\kappa}
\cos{\theta}-\frac{A}{\sqrt{2}} \sin {\kappa} \sin {\theta} \right).
\end{align}

We choose $\Phi$ as eigenstate of $\sigma_{\theta x}$ since the coefficient
$\Delta_B\sin \theta$  in eq.(\ref{cal-E}) is an odd function of
$\kappa$. Then terms with $\sigma_{\theta y}$ and $\sigma_{\theta z}$ must
combine to form the annihilator. Furthermore, since the coefficient
of $\sigma_{\theta x}$ must be real, and be an odd function of
$\kappa$, we require $\gamma_{\theta x}=0$.
This condition determines
$\theta$ in terms of $\kappa$ as
\begin{align}
\sin{\theta}
&=
-\frac{\tan{\kappa}}{\sqrt{\tan^2{\kappa}+8/A^2}},
\label{sin-mu}\\
\cos{\theta} & =
\frac{\sqrt{8}/A}{\sqrt{\tan^2{\kappa}+8/A^2}}.
\label{cos-mu}
\end{align}
In this way,
we decompose $P_{11} = H_{11}+F_{11}$ in
eq.(\ref{TB112}) as follows:
\begin{align}
H_{11} & = \Delta_B \sin{\theta} \, \sigma_{\theta x},\\
F_{11} &= (\Delta_B \cos{\theta}+\gamma_{\theta z} )\sigma_{\theta
z}+\gamma_{\theta y} \sigma_{\theta y}. \label{lambda}
\end{align}
By diagonalizing $H_{11}$,
the eigenenergy $E_\uparrow (\kappa)$ is derived as
\begin{align}
E_\uparrow (\kappa) =s\Delta_B \sin{\theta},\quad (s=\pm 1),
\label{E-kappa}
\end{align}
where only one of the signs $\pm$ is relevant, as derived shortly.
Note that the spectrum
has a form analogous to the case of straight edge given by eq.(\ref{E01}).
The condition for $F_{11}$ to form the annihilator is given by
\begin{align}
\Delta_B \cos{\theta}+\gamma_{\theta z} =  is\gamma_{\theta y}
\label{annihi}
\end{align}
which determines $\lambda$ for the edge mode
as
\begin{align}
\lambda_{\pm}(s)  = \frac 1{2\cos \kappa}\cdot \frac{-\Delta_B
\cos{\theta} \pm \sqrt{R}} { 2/\cos{\theta}+s A/\sqrt{2} },
\label{lambda_pm}
\end{align}
with
\begin{align}
R =
\Delta_B^2
\cos^2{\theta}-2\cos^2{\kappa}
\left(8\cos^{-2}{\theta}-A^2
\right).
\end{align}
In the case of $R<0$,
we obtain complex $\lambda_\pm(s)$
with absolute value \\
\begin{equation}
|\lambda_{\pm}(s)|^{2}=\frac{\sqrt{8}-s A\cos{\theta}}{\sqrt{8}+s
A\cos{\theta}} , \label{absolute-value}
\end{equation}
which is less than unity only for $s=1$, and
positive for $A^2 \le 8$.
Therefore, the edge mode must have $s=1$ in eq.(\ref{E-kappa}), and
the group velocity is positive (negative) in TP-1 (TP-2) regime.
A special case occurs with $\theta=\pm \pi/2$ that corresponds to $\kappa=\mp \pi/2$ according to eqs.(\ref{sin-mu}) and (\ref{cos-mu}).
Actually eq.(\ref{lambda_pm}) gives
$\lambda_{\pm}=\pm i$
at $\kappa=\pi/2$.
Thus, at the boundary of the 1D BZ, there are no edge modes since $|\lambda_{\pm}|=1$.
The neighborhood of this special point has $|\lambda_\pm|<1$, and there
should be an edge mode.
We emphasize that this property is independent of $\Delta$ and $A$, and is specific to the zigzag edge.

\begin{figure}
\begin{center}
\includegraphics[width=0.8\linewidth]{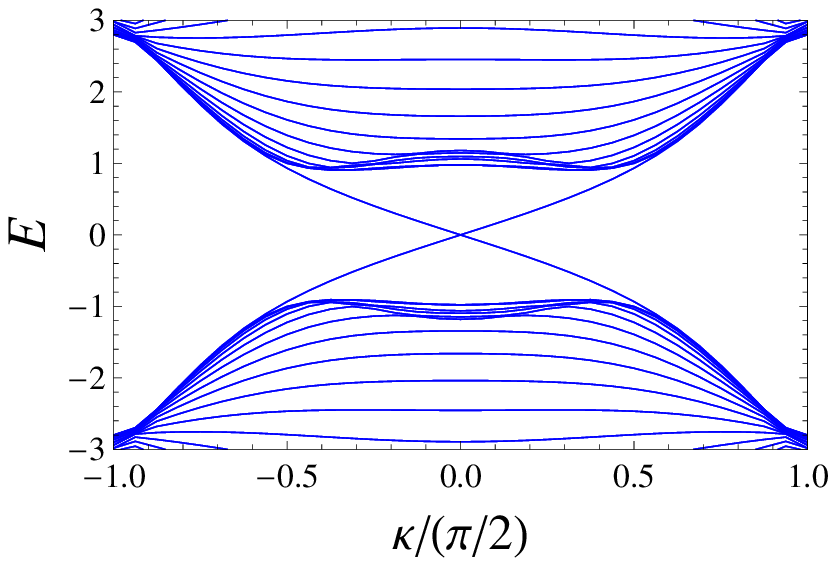}\\
\includegraphics[width=0.8\linewidth]{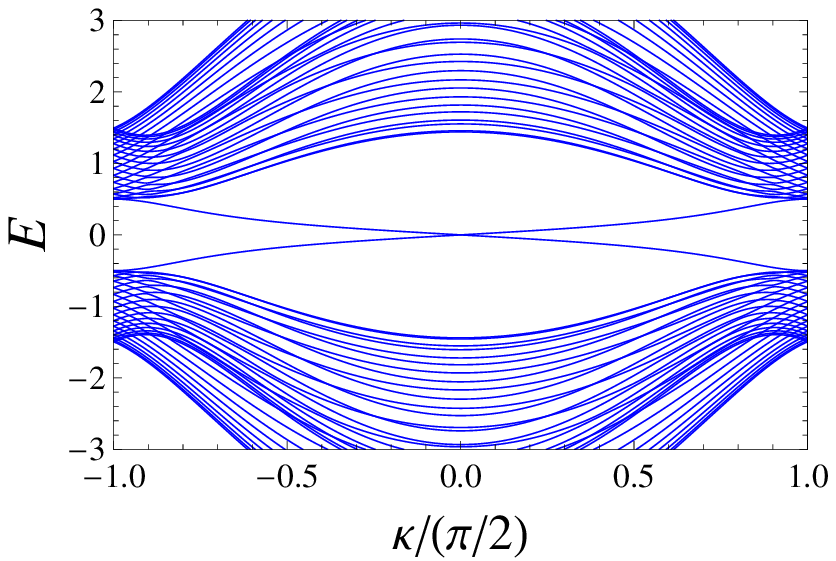}\\
\includegraphics[width=0.8\linewidth]{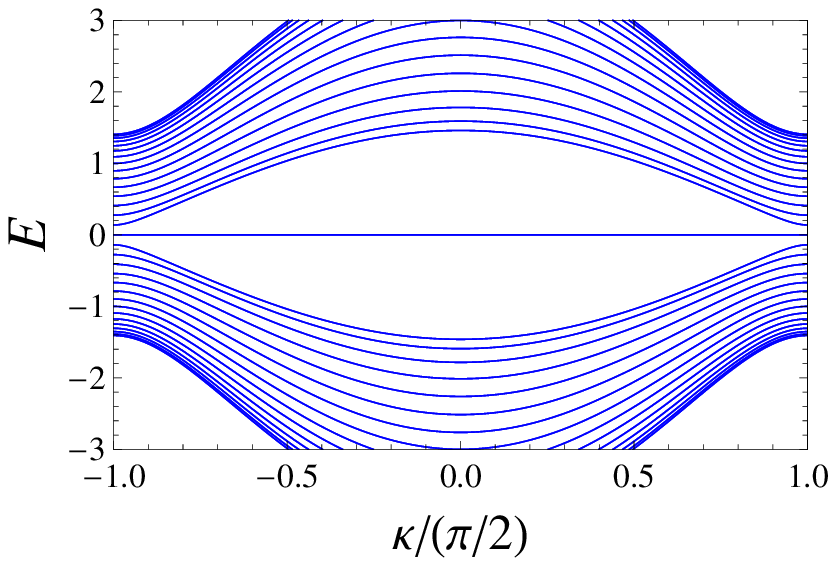}
\caption{Spectrum of zigzag edge modes.  The top panel is for TI-1 state with
$\Delta=1.2$, the center is for TI-2 state with
$\Delta=4.5$.
The bottom panel shows the boundary case $\Delta=4$.
All cases have $A=1$.}
\label{fig4}
\end{center}
\end{figure}

Due to the time-reversal symmetry, we obtain the Kramers partner from
$h_\downarrow$ with the
spectrum $E_\downarrow(\kappa)=- \Delta_B \sin {\theta}$.
Figure \ref{fig4} shows the edge modes together with bulk excitations for the TI-1 and TI-2, and the boundary case $\Delta_B=0$.
The bulk spectrum illustrated is obtained from
$E_{\rm b}(\kappa,\xi)$ as a function of $\kappa$ with fixed $\xi$.
In both regimes,
the edge mode becomes gapless at $\kappa=0$.
Hence only
$\kappa=0$ is the relevant point
where a pair of edge modes are degenerate by time-reversal invariance.
We note that
the spectrum becomes completely flat with $\Delta_B=0$ \cite{imura}.

In the low momentum
region, the spectrum tends to the
linear dispersion
\begin{align}
E_
{\uparrow,\downarrow} (\kappa)= \mp\frac{1}{\sqrt{8}} A \Delta_B\kappa
= \mp\frac{1}{4} A \Delta_B p.
\end{align}
Especially,
with $\Delta_B=\pm 4$,
the spectrum becomes the same as the corresponding modes in the straight edge.
Thus we find that the system acquires the axial symmetry
only in the case of $\Delta_B=\pm 4$ even in the long-wavelength limit.
This is not surprising since the difference in the edge shape remains
even for long wavelength.

\subsection{Allowed momentum range for edge modes}
Let us first 
consider the edge mode in TI-1 regime
corresponding to right-going edge mode (see Figure \ref{fig2} top panel), and $\lambda_{\pm}=\lambda_{\pm}(1)$.
We restrict to the region of positive $\kappa$.
Since complex $\lambda$ always has a pair of solutions with
$|\lambda_{\pm}|\le 1$,
the merging momentum
$\kappa_m <\pi/2$ can be obtained
from the condition
\begin{equation}
\lambda_{+} =1,
\label{lambda+}
\end{equation}
which is equivalent to
\begin{equation}
\left(8-A^2\right)\cos^2{\kappa_m}+2\Delta_B \cos {\kappa_m}+A^2 =0,
\end{equation}
according to eq.(\ref{lambda_pm}).
Note that there is no solution for $0\leq \cos \kappa_m\leq 1$ in the case of $A^2 \geq 8$.
The relevant solution in the case of $A^2< 8$
is given by
\begin{equation}
\cos {\kappa_{m}} =\frac{-\Delta_B \pm \sqrt{\Delta_B^2-A^2 \left(8-A^2
\right)}}{8-A^2},
\label{kappa_m}
\end{equation}
The critical value of $\Delta_B$ beyond which no real $\kappa_m$ exists is given by
\begin{equation}
\Delta_{4c}=-A\sqrt{8-A^2} <0,
\label{real}
\end{equation}
where we consider only the case $0<A<2$ as in the 
the straight edge.
According to eq.(\ref{real}), we have three cases:\\
(i) no solution for $\kappa_{m}<\pi/2$ with
$\Delta_B>\Delta_{4c}$;\\
(ii) single $\kappa_{m}$ with  $\Delta_B=\Delta_{4c}$; \\
(iii) two solutions $\kappa_{m1} < \kappa_{m2}$ with
$\Delta_B < \Delta_{4c}$.\\
Edge modes in the case of $A>2$ will be discussed in \S 5.

The threshold of bulk excitations can occur either at
$\xi=0$ or $\xi \neq 0$ depending on $\kappa$.
The critical value $\kappa_{c}$ separating the two cases
is determined by the condition
\begin{align}
\partial^2 E_{\rm b}(\kappa_c, \xi)/\partial \xi^2\vert _{\xi=0} =0,
\end{align}
which can be reduced to
\begin{align}
2(A^2-4)\cos^2 \kappa_{c}-2\Delta_B\cos \kappa_{c}-A^2=0.
\end{align}
Then we obtain
\begin{align}
\cos \kappa_{c} =\frac{\Delta_B\pm\sqrt{\Delta_B^2+2A^2(A^2-4)}}{2(A^2-4)}.
\label{kappa_c}
\end{align}
In the case of $\xi \neq 0$,
the  momentum $\xi$ at the threshold satisfy the condition
\begin{align}
&\cos \xi =\frac{-2\Delta_B\cos \kappa}{4+(4-A^2)\cos 2\kappa }.
\label{nu-min}
\end{align}
At the zone boundary $\kappa=\pi/2$, eq.(\ref{nu-min}) gives $\xi=\pm\pi/2$ as the solution.
Here the edge mode has the energy
$E_{\uparrow}(\pi/2)=-\Delta_B$,
and the lowest bulk excitation has the same energy
$E_{\rm b}(\pi/2,\pm\pi/2)=-\Delta_B$.
Namely,
$\kappa=\pi/2$
is always a merging point in the zigzag edge
for any parameter settings.

In the following, we analyze the spectrum of the edge mode near the merging momentum according to classification (i), (ii), (iii) given above.

\subsubsection*{Edge modes for the whole BZ}

\begin{figure}
\begin{center}
\includegraphics[width=0.8\linewidth]{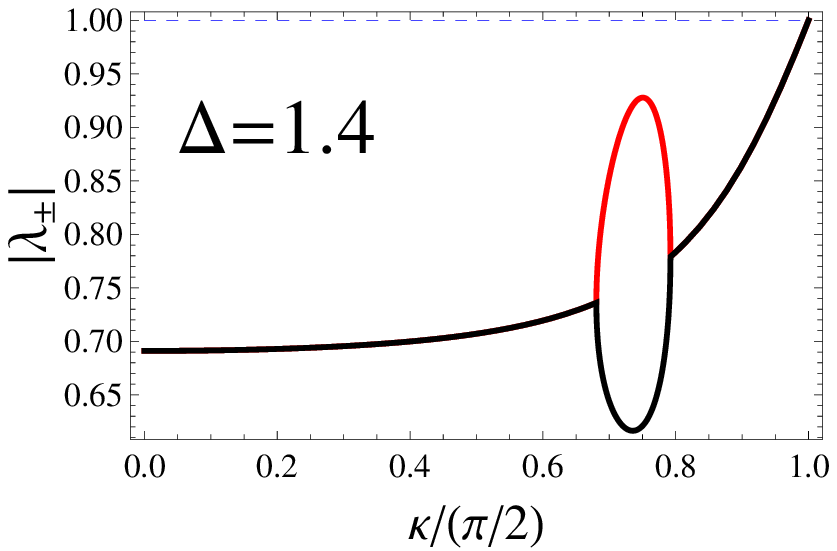}\\
\includegraphics[width=0.8\linewidth]{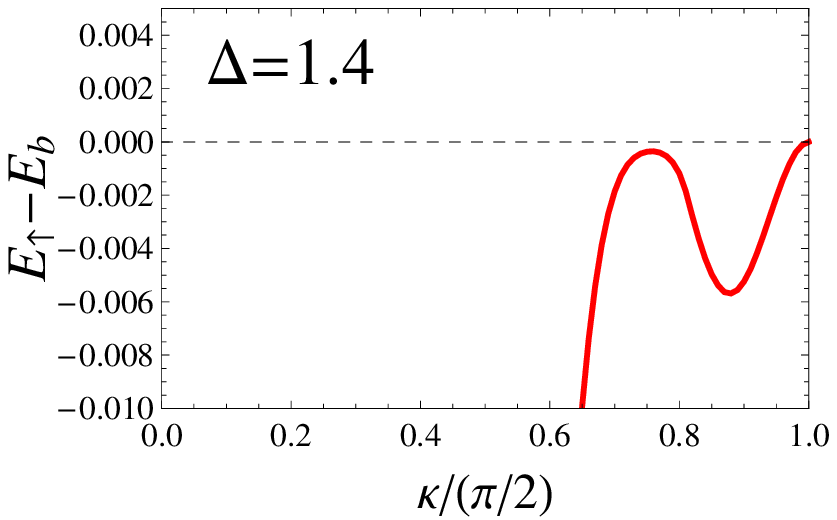}
\caption{
The parameter
$|\lambda_{\pm}|$ and energy difference 
$E_{\uparrow}-E_{\rm b}$ between edge mode and the threshold of bulk excitations as a function of $\kappa$ for  $ \Delta=1.4$ with $A=1$.
For complex $\lambda$,  we obtain $|\lambda_+| = |\lambda_-|$.
In the region of $\kappa$ with real $\lambda_\pm$, $E_{\uparrow}-E_{\rm b}$ becomes almost zero, but is marginally negative.
}
\label{no-km}
\end{center}
\end{figure}

Let us first consider the case (i): $\Delta_B>\Delta_{4c}$.
With $A=1$, we obtain $\Delta_{4c}=-\sqrt 7$, {\it i.e.}, $\Delta_c\sim 1.354$ from eq.(\ref{real}).
Figure \ref{no-km} shows $|\lambda_\pm|$ and the energy of the edge mode
relative to the threshold of bulk excitations in this case with $\Delta=1.4$.

At the zone boundary, the edge mode merges with bulk excitations.
The difference of energies is expanded as
\begin{align}
E_{\uparrow}(\kappa)-E_{\rm b}(\kappa,\frac \pi 2) =  \frac{A^{2}}{\Delta_B}(\kappa-\frac{\pi}{2})^{2}+O\left(
(\kappa-\frac{\pi}{2})^{4} \right).
\label{E1-Eb}
\end{align}
Note that only even order terms appear in the expansion since both
the lowest bulk excitation $E_{\rm b}$ and edge mode energy $E_{\uparrow}$
are symmetric around $\pi/2$.

\subsubsection*{Edge modes with critical momentum $\kappa_m$}

\begin{figure}
\begin{center}
\includegraphics[width=0.8\linewidth]{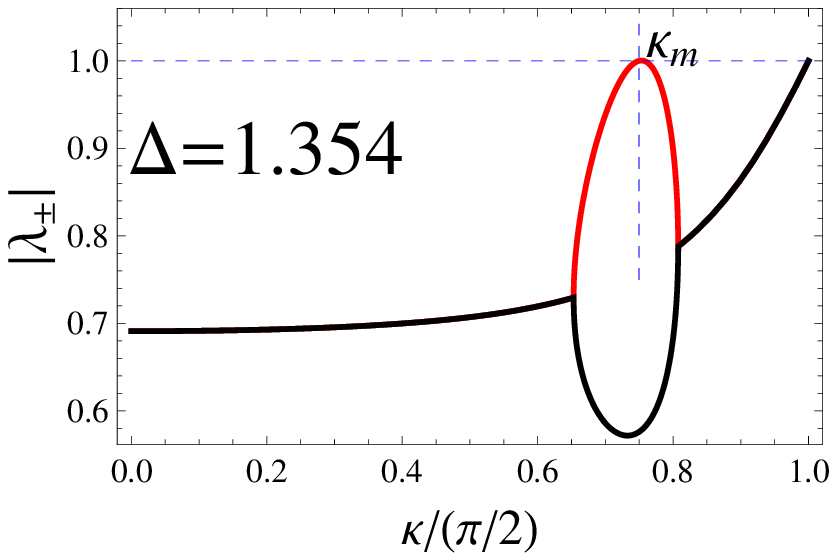}\\
\includegraphics[width=0.85\linewidth]{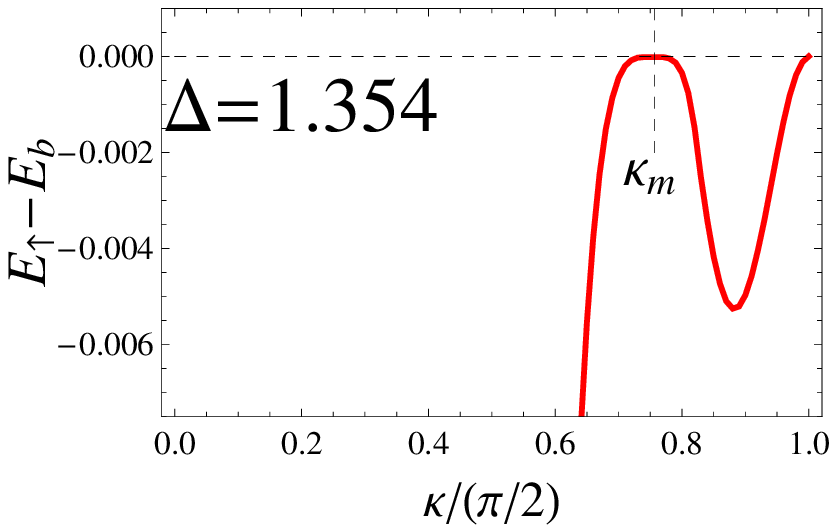}\\
\end{center}
\caption{The same quantities as in Fig.\ref{no-km}
but with $ \Delta=\Delta_{c}=4-\sqrt{7} \sim 1.354$.
Note the scale of the ordinate in the lower panel, showing the minute energy difference as compared with the overall energy scale of the system.
}
\label{critical}
\end{figure}

Next we consider
the critical case (ii) characterized by single $\kappa_{m}$ with  $\Delta_B=\Delta_{4c}$.
Figure \ref{critical} shows $|\lambda_\pm|$ and the energy difference.
By comparing eqs.(\ref{kappa_m}) and (\ref{kappa_c}), we obtain
\begin{align}
0<\kappa_{c1}< \kappa_{m} <\kappa_{c2} < \pi/2.
\end{align}
Hence we find $\xi=0$ for the bulk momentum at the merging point.

\subsubsection*{Edge modes with reentrance}

We finally consider the case (iii): two solutions $\kappa_{m1} < \kappa_{m2}$ with
$\Delta_B < \Delta_{4c}$.
From eqs.(\ref{kappa_m}) and (\ref{kappa_c}) we obtain the relation
\begin{align}
0< \kappa_{c1}< \kappa_{m1}< \kappa_{m2}<\kappa_{c2}< \pi/2
\end{align}
Hence we obtain the bulk momentum $\xi=0$ for both merging points.
Then we expand
$E_{\uparrow}-E_{\rm b}$ around merging points
$\kappa_{mi}\ (i=1,2)$ as follows:
\begin{align}
E_{\uparrow}(\kappa)-E_{\rm b}(\kappa,0) = a_{i}(\kappa-\kappa_{mi})^2+O\left(
(\kappa-\kappa_{mi})^{3} \right),
\label{E-expansion}
\end{align}
where the expansion coefficient is calculated as
\begin{align}
a_{i}&=\frac{4(A^{4}-8A^{2}+\Delta_B^2)}{A^{2}\Delta_B} \nonumber\\
& \times
\left[
\frac{A^{2}-4}{A^2-8}+(-1)^{i}\frac{4\sqrt{A^{4}-8A^{2}+\Delta_B^{2}}}{\Delta_{4}(A^2-8)}
 \right]^{1/2}
\end{align}
Hence we have proven that the threshold of bulk excitation shares the same
energy and velocity with edge mode, since the lowest order term of
expansion is of second order.

At critical value of $\Delta_B=\Delta_{4c}$, we obtain
$\kappa_m=\kappa_{m1}=\kappa_{m2}$,
and both second and third order terms in eq.(\ref{E-expansion})
tend to zero.
Then the
expansion around $\kappa_{m}$ begins from fourth order.
This explains the nearly flat shape of $E_\uparrow-E_{\rm b}$ in Fig.\ref{critical}.

\begin{figure}
\begin{center}
\includegraphics[width=0.8\linewidth]{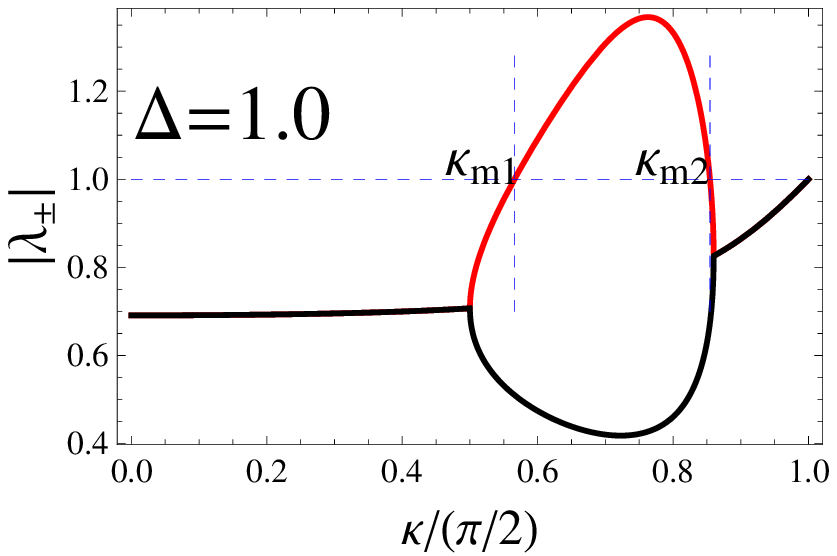}\\
\includegraphics[width=0.85\linewidth]{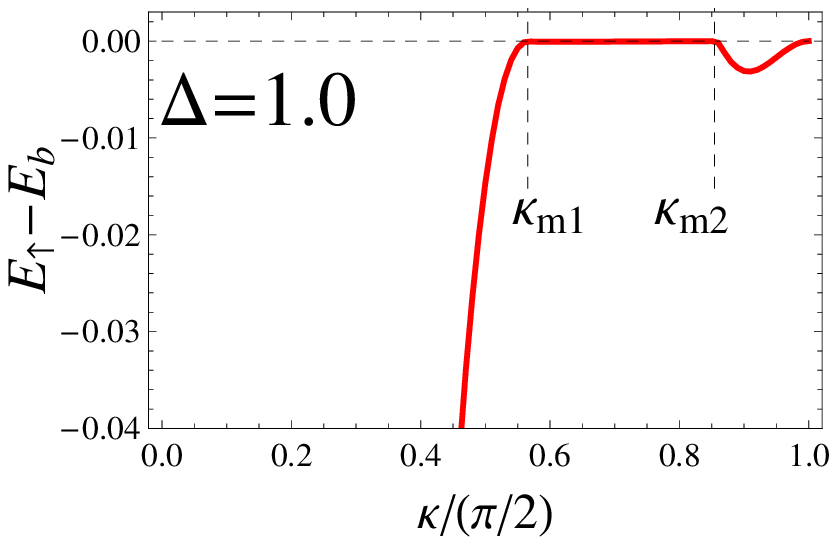}\\
\end{center}
\caption{The same quantities as in Fig.\ref{no-km}
but with $\Delta=1$.
There is no edge mode for $\kappa$ between $\kappa_{m1}$ and $\kappa_{m2}$
where one of $|\lambda_\pm|$ exceeds unity.
}
\label{fig5}
\end{figure}

Presence of two merging points causes novel phenomenon in zigzag edge
mode.
Figure \ref{fig5} shows the parameter $|\lambda_\pm|$ and the energy difference as in previous cases.
In the region $0<\kappa<\kappa_{m1}$ and $\kappa_{m2}<\kappa<\frac{\pi}{2}$,
we obtain $|\lambda_{\pm}|<1$ and
$E_{\uparrow}-E_{\rm b}<0$.
Namely, the edge mode has two
separate momentum ranges of its existence.
The normal edge state (I) starts from zone center
($\kappa=0$), and vanishes at an intermediate point
$\kappa=\kappa_{m1}$.
In addition the reentrant part (II) appears near the zone
boundary: $\kappa_{m2}<\kappa<\frac{\pi}{2}$.
The reentrant edge mode has only marginally
lower energy than lowest bulk excitation, which approximatively
obeys $E_{\uparrow}-E_{\rm b}\propto (\kappa-\kappa_{m2})^{2}$.
In the momentum range
$\kappa_{m1}\leq
\kappa\leq \kappa_{m2}$, the
edge mode disappears since one of $|\lambda_{\pm}|$ exceeds unity.

Let us summarize the results for different $\Delta$ with fixed $A$ shown in
Figs.~\ref{no-km}, \ref{critical} and \ref{fig5}.
As $\Delta$ increases from zero,
the two separate regions for the edge modes widen in momentum space simultaneously.
Namely, both
$\kappa_{m1}$ and $\pi/2-\kappa_{m2}$
increase, while $\kappa_{m2}-\kappa_{m1}$ decreases with
increasing $\Delta$.
At critical $\Delta_{c}$, the two regions merge
($\kappa_{m1}=\kappa_{m2}$), and
the unbroken edge mode emerges that disappears only at the zone boundary.

\subsection{Edge modes in TI-2 regime}

Let us now consider the TI-2 regime with $\Delta_B>0$.
Since the Hamiltonian $H(\mib k)$ has only a linear term of $\Delta_B$, the solution in the TI-2 regime can be obtained from that in the TI-1 regime by changing the sign of $\Delta_B$.
This conversion was indeed made in \S 3 for the straight edge.

In the zigzag edge, the spectrum given by eq.(\ref{E-kappa}) has a negative slope because of $\Delta_B>0$.  
Corresponding to eq.(\ref{lambda-TI-1}),
the merging momentum $\kappa_m>0$ is obtained from the condition
\begin{align}
\lambda_- =-1,
\end{align}
instead of eq.(\ref{lambda+}).
Then we obtain
\begin{equation}
\cos {\kappa_{m}} =\frac{\Delta_B \pm \sqrt{\Delta_B^2-A^2 \left(8-A^2
\right)}}{8-A^2},
\end{equation}
instead of eq.(\ref{kappa_m}), and 
\begin{equation}
\Delta_{4c}=A\sqrt{8-A^2} >0,
\end{equation}
instead of eq.(\ref{real}).
It is clear that the resultant solution of $\kappa_m$ is the same as that in TI-1 range with the same $|\Delta_B|$.  Similarly, one can check that all relevant quantities such as 
$\kappa_{m1}, \kappa_{m2}, 
\kappa_{c1}, \kappa_{c2}$
also have the same correspondence. 
The expression given by eqs.(\ref{E1-Eb}) and (\ref{E-expansion}) remains valid, which means that the difference is now positive.
This is naturally understood since the edge mode $E_\uparrow (\kappa)$ has the negative slope.


\section{Summary and discussion}

We have analytically obtained spectrum and wave function of helical
edge states for the BHZ model by identifying the annihilator for each case of (1,0) and (1,1) edges.
The simplicity of the BHZ model has allowed us to obtain the complete information of the edge modes.
Let us finally consider the case $A>2$.
In the straight edge, 
$\lambda_\pm$ are always real in this case.
Except for this difference, the property of the edge mode spectrum remains the same.
In the zigzag edge, the case
$A>\sqrt 8$ allows no solution for merging momentum $\kappa_m$.  This means the edge mode is present up to the zone boundary.

Edge spectrum shows different properties depending on edge geometry.
In (1,0)-edge case, edge spectrum is proportional to $\sin k$. As
the sign of $\Delta_B$ changes from negative to positive, the main
location of edge mode moves from the center of the 1D BZ to the
boundary. This movement is associated with the change of location of
the bulk energy gap.

For the (1,1) edge, we have obtained the spectrum in the form of
$\pm \Delta_B\sin {\theta(\kappa)}$,  where $\theta(\kappa)$ is an odd
function of momentum $\kappa$. Since $\theta (\kappa)$ does not depend
on $\Delta_B$ as shown in eqs.(\ref{sin-mu}) and (\ref{cos-mu}),
$\Delta_B$ appears only as the scale factor in the spectrum. With
$\Delta_B=0$, the edge modes become completely flat for the whole
Brillouin zone, and two-fold degenerate as a consequence of the
time-reversal symmetry.

The edge mode in the (1,1) geometry contains a novel
reentrant part with extremely small binding energy.
As seen from Fig.\ref{fig5},
the binding energy for the reentrant part is only $10^{-3}$ of the overall energy.
Except for many-body phenomena such as superconductivity and Kondo effect, we have been unaware of emergence of such extraordinary different energy scale.
In spite of the tiny binding energy, the decay of the wave function toward inside the system looks quite normal, as judged by $|\lambda_\pm|$ which deviates clearly from unity.

Mathematically speaking, the tiny binding energy stems from the following factors:\\
(i) The zone boundary $\kappa=\pi/2$ is a special point where the edge mode merges with bulk excitations.\\
(ii) Group velocity of the edge mode is the same as that of threshold excitation in the bulk at the merging point.\\
Let us assume $\kappa\ge 0$ for simplicity.
The condition (ii) requires the energy difference
$E_{\uparrow}-E_{\rm b}$ to be proportional to
$(\kappa-\pi/2)^{2}$ near the zone boundary, and simultaneously to
$(\kappa-\kappa_{m2})^{2}$ near the critical momentum.
Hence in the intervening region
$\kappa_{m2} <\kappa<\pi/2$,
the growth of $E_{\uparrow}-E_{\rm b}$ is constrained from both ends.
Provided $\pi/2-\kappa_{m2}\sim 0.1\pi$, we obtain the scaling factor $\sim 10^{-2}$ from the quadratic dependence of $E_{\uparrow}-E_{\rm b}$.
It is hoped that more physical explanation can be provided in the near future why the binding energy is so small.

\end{document}